\documentclass[aps,pre,twocolumn,superscriptaddress,showpacs,amssymb,amsmath,letterpaper]{revtex4}

\usepackage{graphicx}
\usepackage{dcolumn}
\usepackage{bm}

\begin{document}
\title{Phase transitions induced by complex nonlinear noise\\
in a system of self-propelled agents}

\author{V.~Dossetti}
\email{dossetti@unm.edu}
\affiliation{Consortium of the Americas for Interdisciplinary Science and 
Department\\of Physics and Astronomy, University of New Mexico, Albuquerque, NM 87131, USA}

\author{F.J.~Sevilla}
\email{fjsevilla@fisica.unam.mx}
\altaffiliation[Present address: ]{Instituto de F\'{\i}sica, UNAM, Apdo. Postal 20-364, 01000 M\'{e}xico, D.F., M\'{e}xico.}
\affiliation{Consortium of the Americas for Interdisciplinary Science and
Department\\of Physics and Astronomy, University of New Mexico, Albuquerque, NM 87131, USA}

\author{V.M.~Kenkre}
\email{kenkre@unm.edu}
\affiliation{Consortium of the Americas for Interdisciplinary Science and
Department\\of Physics and Astronomy, University of New Mexico, Albuquerque, NM 87131, USA}

\begin{abstract}
We propose a comprehensive dynamical model for cooperative motion of self-propelled particles, e.g., flocking, by combining well-known elements such as velocity-alignment interactions, spatial interactions, and angular noise into a unified Lagrangian treatment. Noise enters into our model in an especially realistic way: it incorporates correlations, is highly nonlinear, and it leads to a unique collective behavior. Our results show distinct stability regions and an apparent change in the nature of one class of noise-induced phase transitions, with respect to the mean velocity of the group, as the range of the velocity-alignment interaction increases. This phase-transition change comes accompanied with drastic modifications of the microscopic dynamics, from nonintermittent to intermittent. Our results facilitate the understanding of the origin of the phase transitions present in other treatments.
\end{abstract}

\pacs{05.40.-a, 05.45.Xt, 64.60.-i, 87.10.-e}

\maketitle

\section{\label{sec:intro}Introduction}
The study of the collective motion that emerges from short-range interactions in systems of self-propelled particles (SPPs) is of great interest nowadays due to potential applications in physics, engineering, and biology \cite{top04,par99,cou05}. In nature, the formation of bird flocks or animal herds implies the occurrence of a condensation in velocity space and a condensation in position space: the constituents of the group move with velocities similar to one another and additionally, tend to form a spatially contiguous collection. Such behaviors have been classified as velocity matching or alignment, flock centering, and collision avoidance or separation \cite{rey87}. The additional element of \emph{noise} is always present in realistic systems, nonetheless, the way it is introduced may have non-trivial effects in the modeling of flocking phenomena \cite{ald07,pim08}. Models in the literature have concentrated in all or subsets of these ingredients \cite{vic95,gre03,gre04,per08a,czi96,shi96,lev00,erd05,dor06,lee06,rom08}. However, they have not studied the effects of nonlinear correlated noise, which in the model presented in this article appears in a natural way.

The physics of dynamic phase transitions is ideally poised to address cooperative phenomena, hence, the considerable activity in this field in recent times \cite{fed07}. One may argue that flocking of real biological systems in nature occurs in continuous time, that alignment is highly important to the process, and that complex nonlinear noise, even  correlated, is an essential ingredient of realistic systems. This argument is the motivation for our study. Succinctly stated, in this article we study the effects of \emph{alignment} interactions, centering and correlated \emph{angular noise}, in a Lagrangian (trajectory-based) description of flocking, meaning that dynamical equations of motion are employed to compute the trajectories of the individual particles. By doing so, we provide a rich description of the different dynamics of the system, including the origin of intermittent behavior displayed by the mean velocity of the group.

This article is organized as follows. In Sec.\ \ref{sec:model} we introduce the model of our study. We explain how noise enters naturally non-linear and correlated, as we describe how the alignment interaction is implemented. In Sec.\ \ref{sec:disres} we define the quantities we use to characterize the system, and report the results as we discuss the origin of the different dynamical regimes observed. Finally, in Sec.\ \ref{sec:concs} we compare our results in qualitative terms with some other models of interest and report our conclusions.

\section{\label{sec:model}Model}
We study $N$ interacting particles, each having position vector ${\bf x}_i$ and velocity vector ${\bf v}_i=d{\bf x}_{i}/dt$ obeying
\begin{eqnarray}
\lefteqn{m\frac{d{\bf v}_i}{dt}+\frac{K}{N}\sum_{j=1}^{N}({\bf x}_i - {\bf x}_j)}&\nonumber \\
&=\gamma\left[\cos(\psi_{i}+\phi_{i}){\bf i}+\sin(\psi_{i}+\phi_{i}){\bf j}-\dfrac{{\bf v}_i}{v_0}\right],
\label{eq:first2}
\end{eqnarray}
where $m$ and $v_0$ are the mass and the preferred speed of each particle respectively, $K$ is the coupling constant for the real-space condensation interaction, and $\gamma$ is the amplitude of the self-propulsion term with propulsive direction $\psi_i$ and noise $\phi_i$. This term derives from the assumption of constant (resultant) propulsion force of Czir\'{o}k \emph{et al.}\ \cite{czi96}, later generalized by Levine \emph{et al.}\ \cite{lev00} to incorporate in a straightforward manner the alignment interaction by directing this force in the motion's mean direction $\psi_i$ of the neighbors of a given particle $i$. We modify this term further in order to introduce correlated angular noise. This is achieved by simply adding the stochastic variable $\phi_i(t)$ to $\psi_i$, as shown in Eq.\ (\ref{eq:first2}), much in the spirit of (but not equivalent to) Ref.\ \onlinecite{vic95}.

In order to determine the mean direction $\psi_i$ of the neighbors of a given particle $i$, the alignment interaction is implemented by allowing particle $i$ to interact only with its $\mu$ nearest neighbors at time $t$, thus, defining the set \mbox{$\zeta_i(t)=\{{\bf v}_{i_1}(t), {\bf v}_{i_2}(t), \ldots,{\bf v}_{i_\mu}(t)\}$} (${\bf v}_i \notin \zeta_i$) with the velocity vectors of the neighbors. We identify $\mu$ as the \emph{connectivity} of the particles, and consider it to be equal for all particles. In this way, given that the individual velocities ${\bf v}_i$ have a magnitude $v_i$ and direction $\varphi_i$, the case $\mu = 0$ corresponds to the absence of alignment interactions, and $\psi_i=\varphi_i$, otherwise
\begin{equation}
\psi_{i}\equiv\arctan\left[\frac{v_i\sin\varphi_i + \displaystyle \sum_{{\bf v}_j \in \, \zeta_i(t)} v_j\sin\varphi_j}
{v_i\cos\varphi_i + \displaystyle \sum_{{\bf v}_j \in \, \zeta_i(t)} v_j\cos\varphi_j}\right].
\label{eq:prop2}
\end{equation}

On the other hand, noise introduced through the stochastic variable $\phi_i$, as in Eq.\ (\ref{eq:first2}), is \emph{intrinsic} to the decision mechanism of each one of the self-propelled agents in that a given agent, receives a clear signal from its neighbors, however, it may ``decide'' to move in a different direction \cite{pim08}. For example, this kind of noise may be thought as arising from the limitations of a bird to follow a specific and precise flight-path (due to fatigue, for example) or some other external factors (wind currents, for instance). In our model, noise appears naturally in a highly non-linear form requiring correlations (that decay in time) to be considered for it to have a non-trivial effect \cite{hor84}, and differs fundamentally from previous studies \cite{vic95,gre03,gre04,per08a,czi96,erd05,lee06,rom08,cha08}. In this work, the correlated random sequences $\phi_{i}(t)$, yet independent between different agents, are obtained by correspondingly coupling the stochastic equation
\begin{equation}
\dot{\phi}_i' = -\lambda \phi_i' + \lambda g_i,
\label{eq:noise1}
\end{equation}
to each one of the equations of motion of the group individuals. Here, $g_i(t)$ is a Gaussian white noise with mean value $\langle g_i(t) \rangle = 0$ and autocorrelation function \mbox{$\langle g_i(t) g_j(t') \rangle = 2D\delta_{ij}\delta(t-t')$}. Thus, the driven noise $\phi_i'(t)$ is then exponentially correlated noise (derived from the Orstein-Uhlenbeck processes) with properties $\langle\phi_i'(t)\rangle=0$ and \mbox{$\{\langle \phi_i'(t) \phi_j'(t) \rangle\} = D\lambda\delta_{ij} \hbox{exp}(-\lambda|t-t'|)$}. The curly brackets $\{\ \cdots \}$ denote averaging over the distribution of initial $\phi_0'$ values, which are equivalent for every particle, and taken from the distribution
\begin{equation}
P_{\phi'}(\phi_0') = \frac{1}{\sqrt{2 \pi D \lambda}} \;\; \hbox{exp}\left(  -\frac{{\phi_0'}^2}{2D\lambda} \right).
\label{eq:pdprime}
\end{equation}
Clearly, the correlation time for the colored noise is \mbox{$t_c = \lambda^{-1}$}. Notice that the noise sequence $\phi_i'(t)$ has a Gaussian distribution \cite{fox88}. However, in order to be able to compare our results to some other models of interest, we would like for our correlated angular noise, $\phi_i(t)$, to have an uniform distribution
\begin{equation}
P_{\eta}(\phi) = \left\{ \begin{array}{ll}
\frac{1}{2\eta} & \mbox{if $|\phi| \leq \eta$} \\
0 & \mbox{otherwise}
\end{array} \right.,
\label{eq:peta}
\end{equation}
in the interval $[-\eta,\eta]$, where $\eta$ (running from $0$ to $\pi$) defines the maximum possible value for $|\phi_i|$. By matching the mean squared variance of the distribution $P_{\phi'}$ of Eq.\ (\ref{eq:pdprime}) with that one of the desired distribution $P_{\eta}$ of Eq.\ (\ref{eq:peta}), i.e., $D\lambda = \eta^2/3$, with a further transformation that preserves the area between them, the correlated sequences of angular noise can be obtained from
\begin{equation}
\phi_i(t) = \sqrt{3D\lambda} \;\; \hbox{erf}\left(\frac{\phi_i'(t)}{\sqrt{2D\lambda}}\right),
\label{eq:noise2}
\end{equation}
where $\hbox{erf}(x)$ stands for the \emph{Gauss error function}. For a given (dimensionless) noise intensity $\eta$, and correlation time $t_c$, the properties of our angular noise are
\begin{subequations}
\begin{eqnarray}
&&\langle \phi_i(t) \rangle = 0, \label{eq:mnphi} \\
&&\langle \phi_i(t) \phi_j(t') \rangle = \frac{\eta^2 \delta_{ij}}{3} \;\; \hbox{exp}\left(-\frac{|t-t'|}{t_c}\right),
\label{eq:autophi}
\end{eqnarray}
\end{subequations}
which correspond to its mean value and autocorrelation function respectively.

Throughout the rest of this paper, we will measure speed in units of $v_0$, time in units of $t_0=mv_0/\gamma$ and distance in units of $x_0=mv_0^2/\gamma$. In this way, only three parameters remain independent: the dimensionless coupling parameter $\kappa=mKv_0^2/\gamma^2$, the connectivity $\mu$, and the intensity of the angular noise $\eta$. All numerical results presented here were obtained by integrating the set of Eqs.\ (\ref{eq:first2}) and (\ref{eq:prop2}) with a modified version of the velocity-Verlet algorithm \cite{gro97}.

\begin{figure}
\includegraphics[width=.4\textwidth]{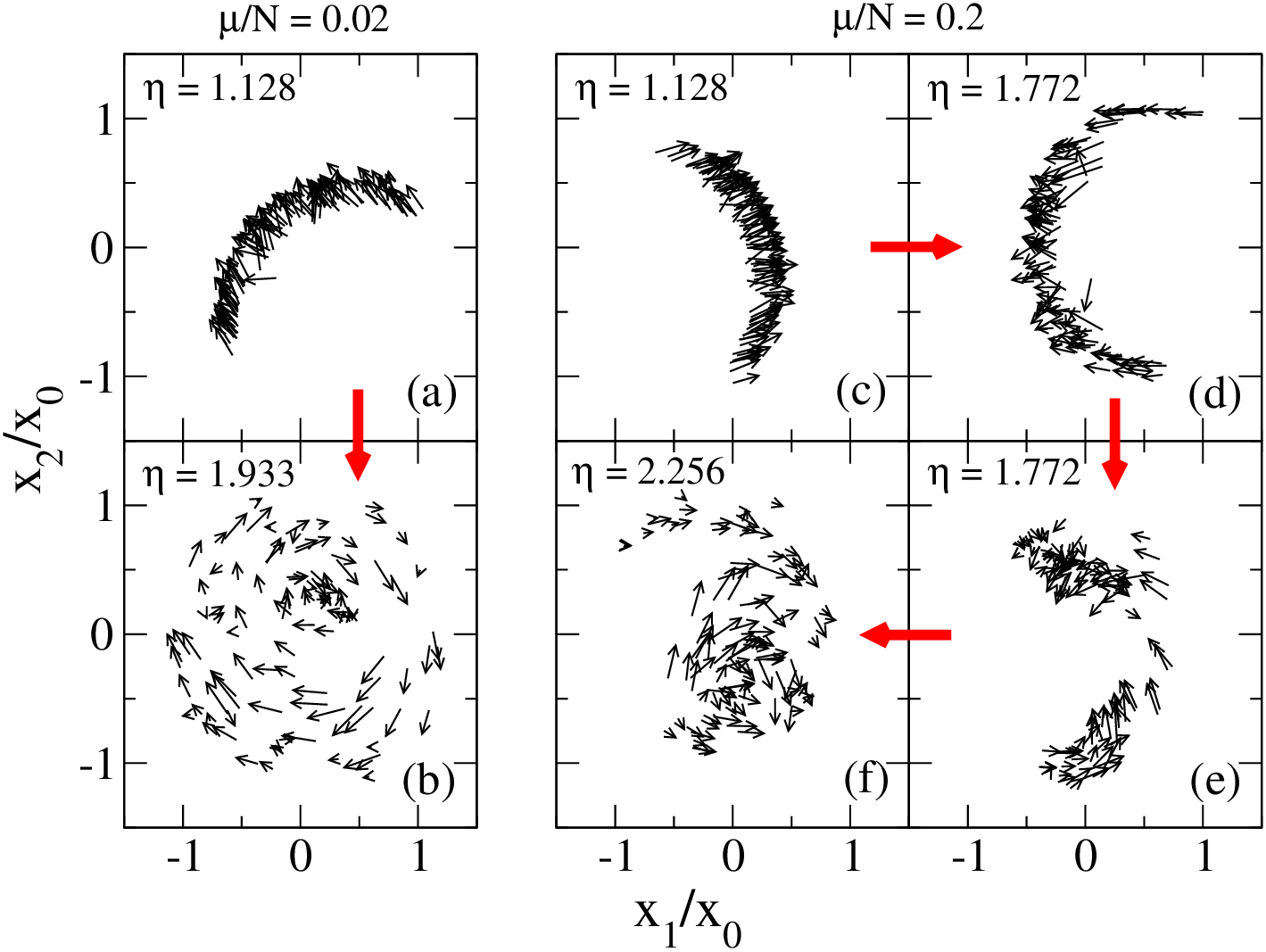}
\caption{(Color online) Snapshots of the configuration in position space for the model given in Eqs.\ (\ref{eq:first2}) and (\ref{eq:prop2}), with $\kappa = 1$ and $t_c=10$, for systems of $N=100$ particles. The small arrows represent the velocity vectors ${\bf v}_i$ for each one of them while the big (red) arrows, across figures, correspond to the way of increasing noise intensity. In (a) and (b), and (c) and (f), the translational and oscillatory states are, respectively, shown in regions I and IV for different connectivity values (see text for more details). When the system shows intermittent dynamics, (d) a laminar regime can be interrupted by (e) a turbulence burst. In (d) and (e), the value of $\eta$ considered was taken inside region II (see text).}
\label{fig:states}
\end{figure}

\section{\label{sec:disres}Discussion and results}
To characterize the system and its stationary states, we monitored quantities such as the instantaneous position and velocity of the center of mass, defined as ${\bf X}_{CM}(t)=\frac{1}{N}\sum_{i=1}^{N}{\bf x}_{i}(t)$ and ${\bf V}_{CM}(t)=\frac{1}{N}\sum_{i=1}^{N}{\bf v}_{i}(t)$, respectively. To quantify the translational and rotational motions of the flock, as order parameters, we measured the normalized mean velocity and angular momentum of the group,
\begin{subequations}
\begin{eqnarray}
\Psi & = & \lim_{T \rightarrow \infty} 
\frac{1}{T} \int_{0}^{T} \frac{1}{v_0} |{\bf V}_{CM}(t)| \, dt,
\label{eq:vcm1} \\
\Lambda & = & \lim_{T\rightarrow\infty} 
\frac{1}{T}\int_{0}^{T} 
\frac{1}{N}\left| \sum_{i=1}^{N} \frac{{\bf L}_i(t)}
{|\tilde{\bf x}_i(t)| |\tilde{\bf v}_i(t)|} \right| \, dt,
\label{eq:normang}
\end{eqnarray}
\end{subequations}
where ${\bf L}_i(t) = \tilde{\bf x}_i(t) \times \tilde{\bf v}_i(t)$ corresponds to the angular momenta of the individual particles with respect to the center of mass, $\tilde{\bf x}_i(t) = {\bf x}_i(t) - {\bf X}_{CM}(t)$, and $\tilde{\bf v}_i(t) = {\bf v}_i(t) - {\bf V}_{CM}(t)$. We also measured quantities regarding the spatial distribution of the particles such as the instantaneous mean square dispersions in the directions parallel ($\Sigma_{||}$) and orthogonal ($\Sigma_{\perp}$) to the direction of ${\bf V}_{CM}$ \cite{erd05},
\begin{subequations}
\label{eq:disps}
\begin{eqnarray}
\Sigma_{||}(t) & = & \frac{1}{N |{\bf V}_{CM}(t)|^2} 
\sum_{i=1}^{N} [\tilde{\bf x}_i(t) \cdot {\bf V}_{CM}(t)]^2, 
\label{eq:spar} \\
\Sigma_{\perp}(t) & = & \frac{1}{N |{\bf V}_{CM}(t)|^2}
\sum_{i=1}^{N} [\tilde{\bf x}_i(t) \times {\bf V}_{CM}(t)]^2.
\label{eq:sperp}
\end{eqnarray}
\end{subequations}

Depending on initial conditions, the values of the normalized connectivity $\mu/N$ and the noise intensity $\eta$, the system may attain a \emph{translational state} (TranS) where all of the particles move more or less in the same direction, shown in Figs.\ \ref{fig:states}(a) and \ref{fig:states}(c), with $\Psi>0$. The curved shape of the group is principally due to correlations in the noise in combination with the centering interaction, for it can be observed even in the absence of any alignment interactions. The system may also attain an \emph{oscillatory state} (OscS) shown in Figs.\ \ref{fig:states}(b) and \ref{fig:states}(f), where particles circle around a center of mass that diffuses with $\Psi\approx0$.

Particular care was taken to obtain initial conditions for the system either in the TranS or the OscS by setting $\eta=0$ at the beginning. Then, in one case, the system was allowed to relax to a stationary TranS under a global alignment interaction condition $\mu/N=N-1$. In the other case, particles were homogeneously placed on a ring with tangent velocity vectors, and the system was allowed to relax to a stationary OscS with $\mu/N=0$. At this point, time is set $t=0$ in all our numerical simulations, and the parameters of the system are correspondingly fixed for each one of the cases studied.

\begin{figure}
\includegraphics[width=.43\textwidth]{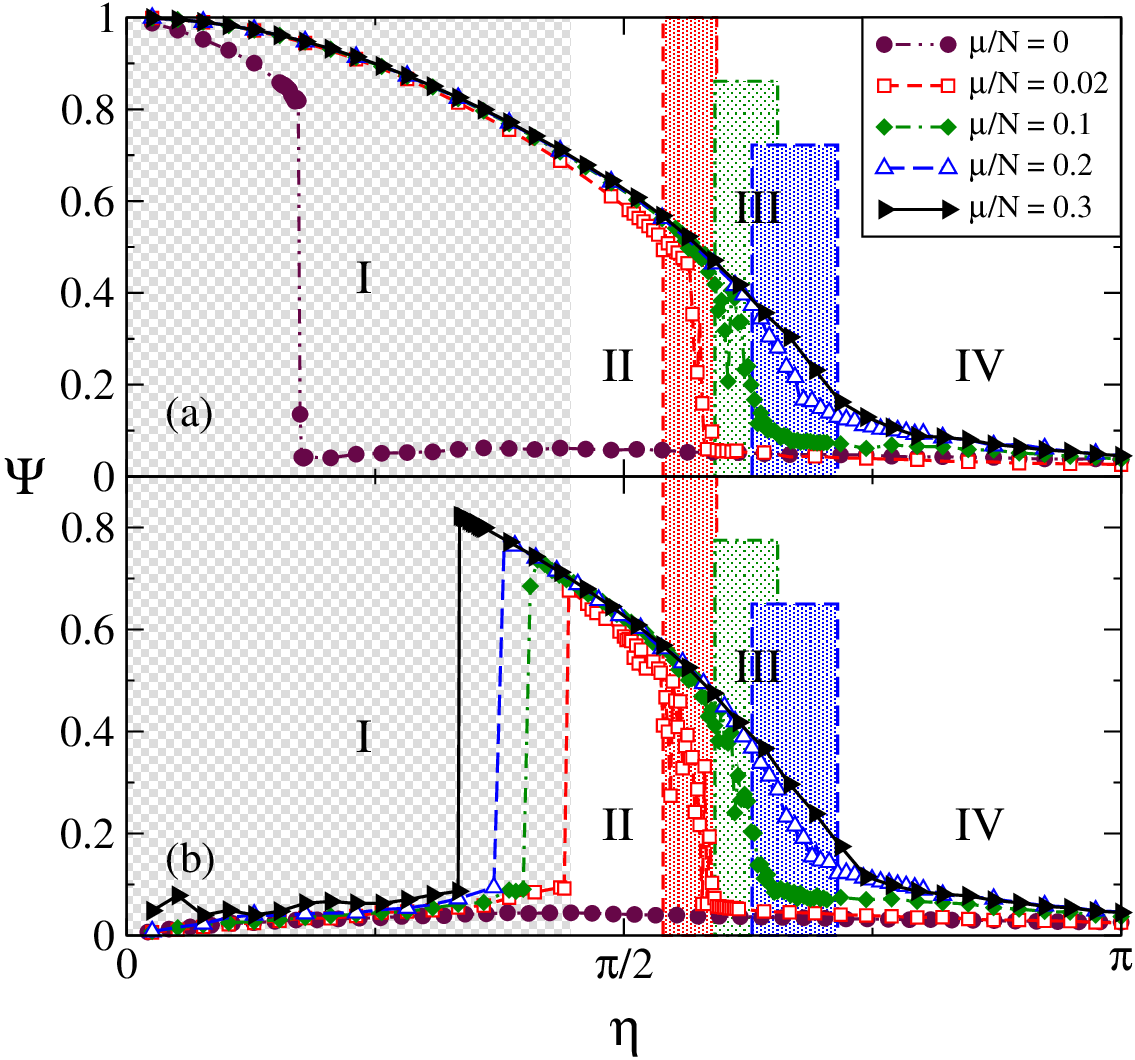}
\caption{(Color online) Stationary order parameter $\Psi$, given in Eq.\ (\ref{eq:vcm1}), as a function of the noise intensity $\eta$. Equations (\ref{eq:first2}) and (\ref{eq:prop2}) were numerically solved for systems of $N=200$ particles with $\kappa=1$ and $t_c=10$. Initial conditions were taken in the TranS (a) and in the OscS (b). The changes of stability and the different kinds of phase transitions are apparent as the connectivity $\mu/N$ changes. See text for more details}
\label{fig:psi}
\end{figure}

\begin{figure}
\includegraphics[width=.4\textwidth]{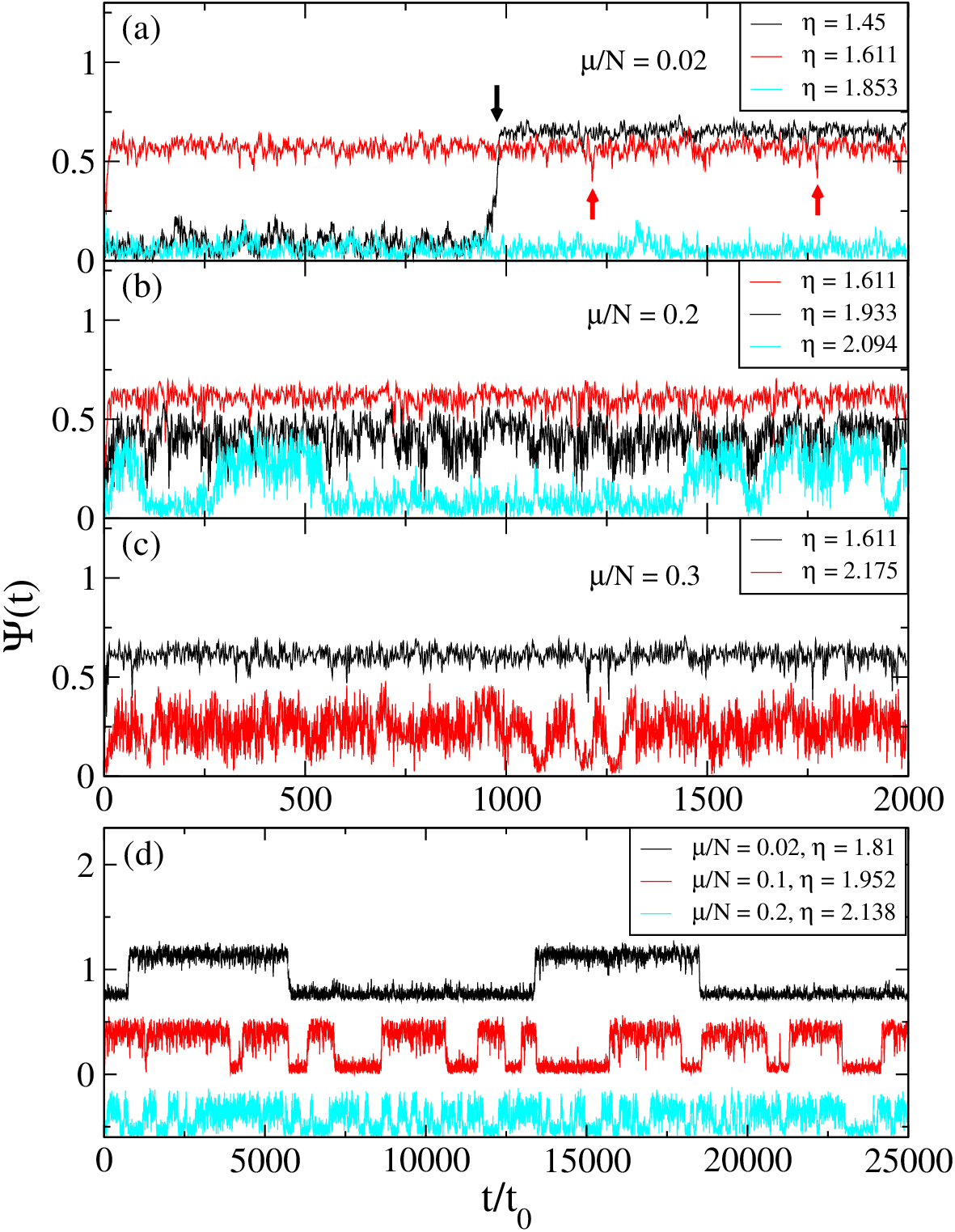}
\caption{(Color online) Times series of the order parameter $\Psi(t)=|{\bf V}_{CM}(t)|/v_0$ for different values $\mu/N$ and $\eta$. Typical cases inside region III for different connectivities are shown in (d), where the curves are vertically displaced for better clarity. Note that the time intervals between turbulence bursts, and between free transitions to the translational and oscillatory states inside reigon III, are in general larger than the time scale of the noise-induced fluctuations for $\mu/N\leq0.2$. See the text for more details.}
\label{fig:psi-tseries}
\end{figure}

\subsection{\label{sec:meanvel}Mean velocity of the group}
In the absence of alignment interactions ($\mu/N=0$), the normalized mean velocity of the group, $\Psi$, undergoes a noise-induced discontinuous phase transition from the TranS to the OscS as shown in Fig.\ \ref{fig:psi}, where plots for $\Psi$ as a function $\eta$ are presented. Both the TranS and the OscS are accessible and stable for subcritical values of the noise, as can be appreciated in Figs.\ \ref{fig:psi}(a) and \ref{fig:psi}(b), where initial conditions were taken in the TranS and the OscS, respectively; this kind of behavior resembles the model of Erdmann \emph{et al.}\ \cite{erd05}. The critical point $\eta_c\approx0.548$ of the phase transition for the case shown in the figure, shifts to lower values as one increases $t_c$ due to stronger effects from the noise. In contrast, increasing the coupling parameter with the center of mass, $\kappa$, enhances the stability of the TranS and $\eta_c$ shifts to higher values. On the other hand, when a global alignment interaction is considered, i.e., when $\mu=N-1$, the TranS solution becomes the only available solution of the system as the phase transition disappears. For this case, it can be analytically estimated that $\Psi\approx\frac{\sin\eta}{\eta}$; we analyze this and some other limit cases in Appendix \ref{sec:limitcases}. Nevertheless, for local alignment interactions with $\mu/N\ll1$, at least four regions with different stability can be identified, depicted with roman numerals in Fig.\ \ref{fig:psi}.

\begin{table}[b]
\caption{Approximate values of the critical points that separate the different stability regions of the cases with finite $\mu/N$ analyzed in Fig.\ \ref{fig:psi}(a).}
\label{tab:cps}
\begin{ruledtabular}
\begin{tabular}{ccccc}
$\mu/N$  & $\eta_{ot}$ & $\eta_{tb}$ & $\eta_{bo}$ & $\eta_{to}$\\ 
& (R.I$\rightarrow$R.II) & (R.II$\rightarrow$R.III) & (R.III$\rightarrow$R.IV) & (R.II$\rightarrow$R.IV)\\ \hline
0.02 & 1.383 & 1.695 & 1.864 & \\
0.1 & 1.253 & 1.817 & 2.056 & \\
0.2 & 1.16 & 1.976 & 2.246 & \\
0.3 & 1.047 & & & 2.385\\
\end{tabular}
\end{ruledtabular}
\end{table}

Region I (gray checked portions in Fig.\ \ref{fig:psi} for the case $\mu/N=0.02$) shows bistability similar to the case $\mu/N=0$, where both states are accessible depending on initial conditions. For larger values of $\eta$, we found a window of noise values (region II) where the TranS is the only available stable solution, whose size increases with the normalized connectivity $\mu/N$. Indeed, the noise-induced fluctuations make the OscS unstable inside this region even if initial condition are taken in this state, consequently driving the system to the TranS [see, for example, the event marked with the downward arrow for the curve with $\eta=1.45$ in Fig.\ \ref{fig:psi-tseries}(a)]. Thus, the system exhibits a discontinuous phase transition in $\Psi$ from region I to region II with critical point $\eta_{ot}$; the values of the critical points for the different cases are presented in Table \ref{tab:cps}.

\begin{figure}
\includegraphics[width=.34\textwidth]{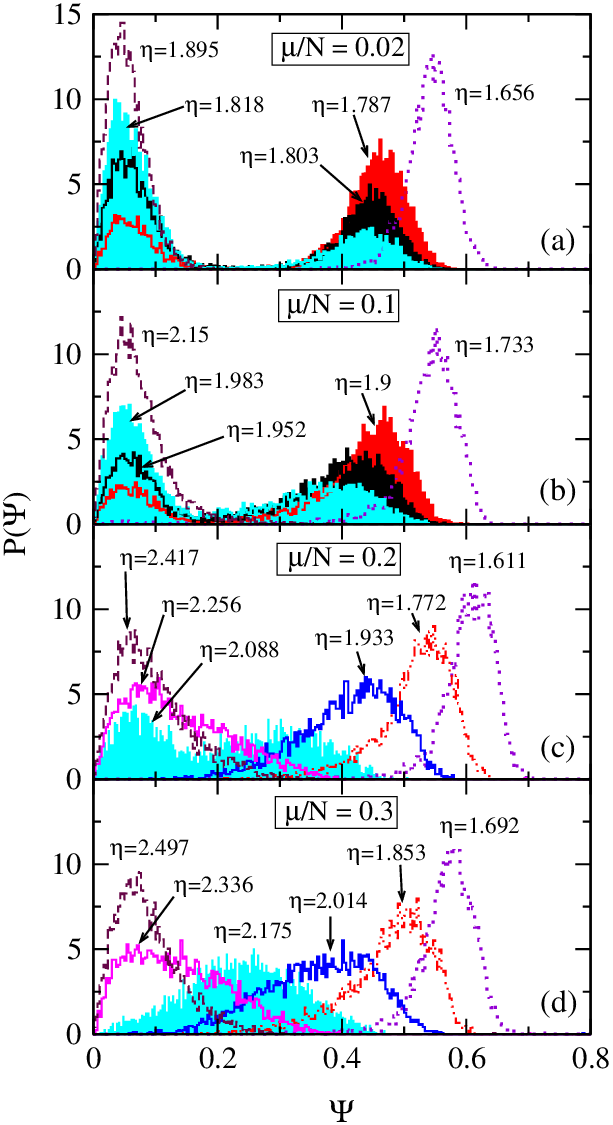}
\caption{(Color online) Stationary probability distribution functions of the order parameter $\Psi$ for selected cases of Fig.\ \ref{fig:psi} with finite $\mu/N$. In (a)-(c), the filled curves correspond to values of $\eta$ inside region III, where the system shows bistability and free transitions between the TranS and the OscS through the bimodal distribution of $P(\Psi)$. This behavior is typical of first-order phase transitions but only close to the critical point. In (d), the filled curve denotes the unimodal character of $P(\Psi)$ in the transition from the TranS to the OscS, typical of second-order phase transitions.}
\label{fig:psi-PDFs}
\end{figure}

Inside region II, the solution for the TranS starts to shows disorder spikes as the system develops intermittent behavior [some of these events are marked with upward arrows for the curve $\eta=1.611$ in Fig.\ \ref{fig:psi-tseries}(a)]. Increasing the connectivity and/or the noise results in an increase in the amplitude of the spikes and, in some cases, the system is able to even segregate into clusters. See, for example, how for a fixed and the same noise intensity, the laminar regime of the TranS shown in Fig.\ \ref{fig:states}(d) is interrupted by a segregation event shown in Fig.\ \ref{fig:states}(e). These events are reminiscent of ``turbulence bursts'' where the structure of the group is directionally inhomogeneous; however, if present, the clusters themselves may show local order [see Fig.\ \ref{fig:states}(e)]. With a further increase in the noise intensity, and for values up to $\mu/N\approx0.2$, a region with mixed dynamics develops in a discontinuous manner (region III), depicted by the shadowed portions in Fig.\ \ref{fig:psi} for the cases $\mu/N=0.02,0.1,0.2$. As can be appreciated, the size of this region increases with the normalized connectivity $\mu/N$.

Inside region III, the stronger turbulence bursts may cause the system to suffer \emph{free transitions} between the TranS to the OscS, in the sense that the system may acquire any of the two states regardless of initial conditions as shown in Fig.\ \ref{fig:psi-tseries}(b) for the curve with $\eta=2.094$. We must mention that turbulence bursts in systems that present intermittent behavior are known to occur at irregular time intervals showing a power-law distribution \cite{cha08,hue04,ber87} and, in our model, not all of the turbulence bursts induce a transition between the TranS and the OscS. What we observed from our numerical results is that, as the connectivity increases, the frequency of these transitions also increases, while the plateaus where the system spends some time in either of the two states become smaller in the mean [see Fig.\ \ref{fig:psi-tseries}(d)]. Moreover, the system spends more time in the TranS at the beginning of region III, while, with the increasing noise, it starts to spend more and more time in the OscS as one approaches the end of this region. This can be better appreciated looking at the probability distribution function (PDF) of $\Psi$ shown in Figs.\ \ref{fig:psi-PDFs}(a) and \ref{fig:psi-PDFs}(b) for $\mu/N=0.02,0.1$. There, the bimodal filled curves correspond to values of $\eta$ inside region III, and corroborate the coexistence of states and the bistability. The rightmost peaks of the distributions correspond to the TranS, and their amplitude diminishes with the increasing noise. The opposite happens to the leftmost peaks that correspond to the OscS, whose amplitude increases with the noise intensity. Nonetheless, this ``transference of stability'' between the translational and oscillatory states occurs in a continuous manner as shown in Figs.\ \ref{fig:psi}(a) and \ref{fig:psi}(b); the strong fluctuations for the cases with $\mu/N=0.02,0.1$ are due to lack of better statistics for the free transitions, requiring longer integration times than those considered in this article.

It is not our intention to give a quantitative detailed analysis of the distribution of the time intervals between the free transitions inside region III here due to the cumbersome computational requirements for these calculations. We will leave this as an open question for future work.

The borders of region III are delimited by the critical points $\eta_{tb}$, for the phase transition that occurs from regions II to III, and $\eta_{bo}$ for the one that occurs from regions III to IV. The values of the critical points for the different cases presented in Fig.\ \ref{fig:psi} are shown in Table \ref{tab:cps}. In particular, the end of region III corresponds to the case when the OscS becomes the only stable solution of the system in region IV, in a contrasting effect from the correlated angular noise in comparison to region II. These transitions from regions II to III, and III to IV, may be considered discontinuous as the coexistence of states inside region III is reflected in the bimodal distribution of $P(\Psi)$ as explained before; see, for example, Figs.\ \ref{fig:psi-PDFs}(a) - \ref{fig:psi-PDFs}(c). Nonetheless, this contrasts to what typical first-order phase transitions show, where the bistability and the coexistence of states can only be observed close to the critical point and not for a continuous window of values of the noise.

\begin{figure}
\includegraphics[width=.43\textwidth]{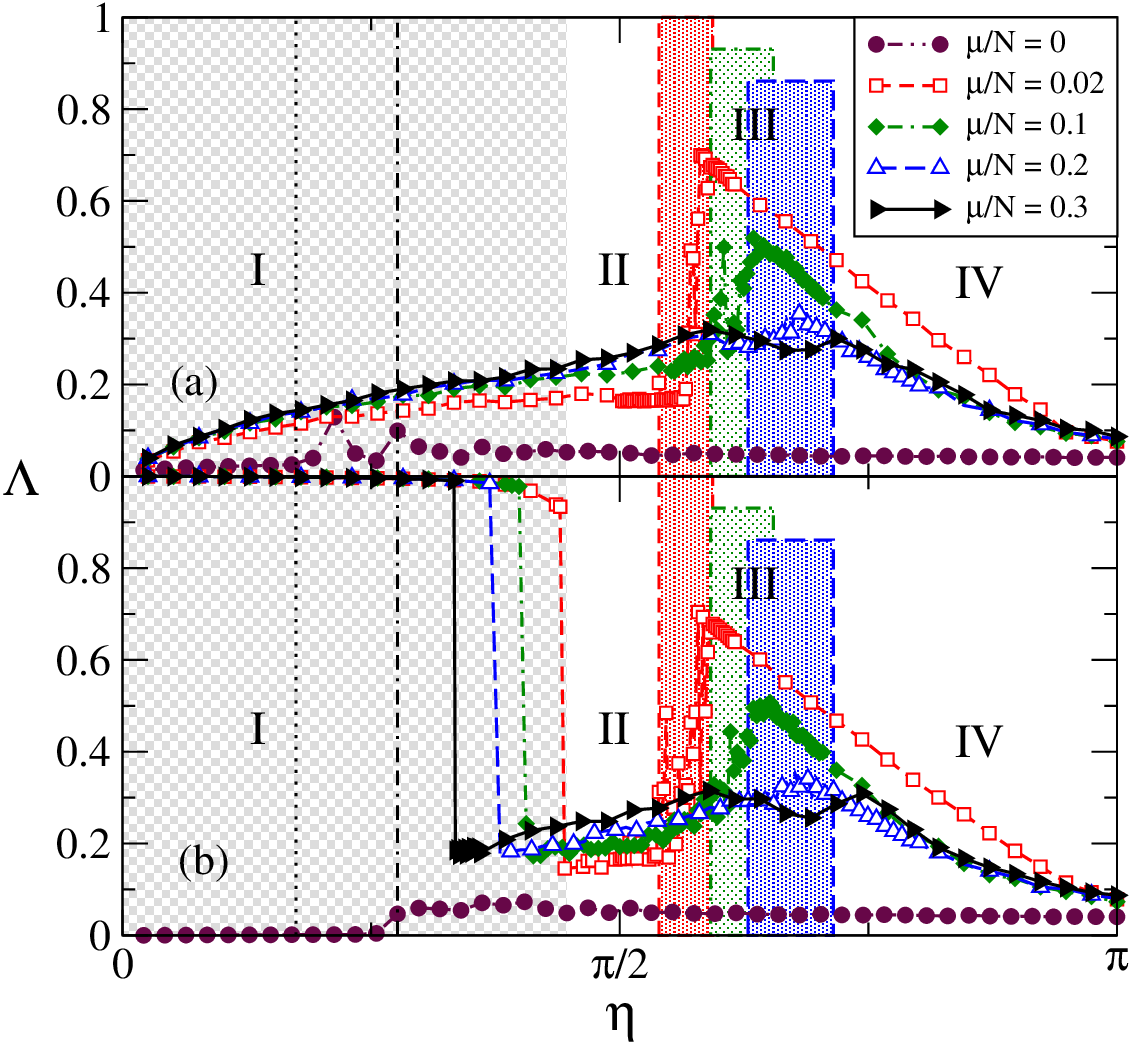}
\caption{(Color online) Stationary angular momentum $\Lambda$, given in Eq.\ (\ref{eq:normang}), as a function of $\eta$ with $N=200$, $\kappa=1$, and $t_c=10$, for initial conditions taken in the TranS (a) and in the OscS (b). For the case $\mu/N=0$, the vertical dotted line corresponds to the critical point $\eta_c$ of the phase transition in $\Psi$ from the TranS to the OscS when initial conditions are taken in the TranS. Beyond that line the only solution of the system is the OscS regardless of initial conditions. On the other hand, beyond the vertical dash-dot-dashed line, the correlated angular noise is strong enough to randomly change the direction of rotation of the particles in the OscS. In contrast, in between the two vertical lines and after the system reaches the stationary OscS, particles rotating in either direction, clockwise or counterclockwise, will remain rotating in such a way regardless of the fluctuations induced by the noise. This is more evident in (a) where $\Lambda$ presents some bumps product of an asymmetric number of particles rotating in both directions, due to the transition the system undergoes from the TranS to the OscS. See text for more details on the cases with finite connectivities.}
\label{fig:mang}
\end{figure}

For connectivities $\mu/N\geq0.3$, a new kind of dynamics develops as the frequency of the occurrence of turbulence bursts and free transitions between the translational and oscillatory states become indistinguishable from the time scale of the fluctuations induced by noise. For this case, $\Psi$ smoothes out completely and, as region III disappears, the transition from the TranS in region II to the OscS in region IV shows an apparent continuous nature with critical point $\eta_{to}$ (see Fig.\ \ref{fig:psi} and Table \ref{tab:cps}). The nature of this phase transition can be inferred from the one-peaked form of $P(\Psi)$ at criticality shown in Fig.\ \ref{fig:psi-PDFs}(d), and from the fact that $\Psi(t)$ does not show clear plateaus that imply the coexistence of states but only strong fluctuations [see the curve for $\eta=2.175$ in Fig.\ \ref{fig:psi-tseries}(c)] as the system goes from the TranS to the OscS. Looking at the angular momentum of the group can provide a further insight (see next subsection).

\begin{figure}
\includegraphics[width=.33\textwidth]{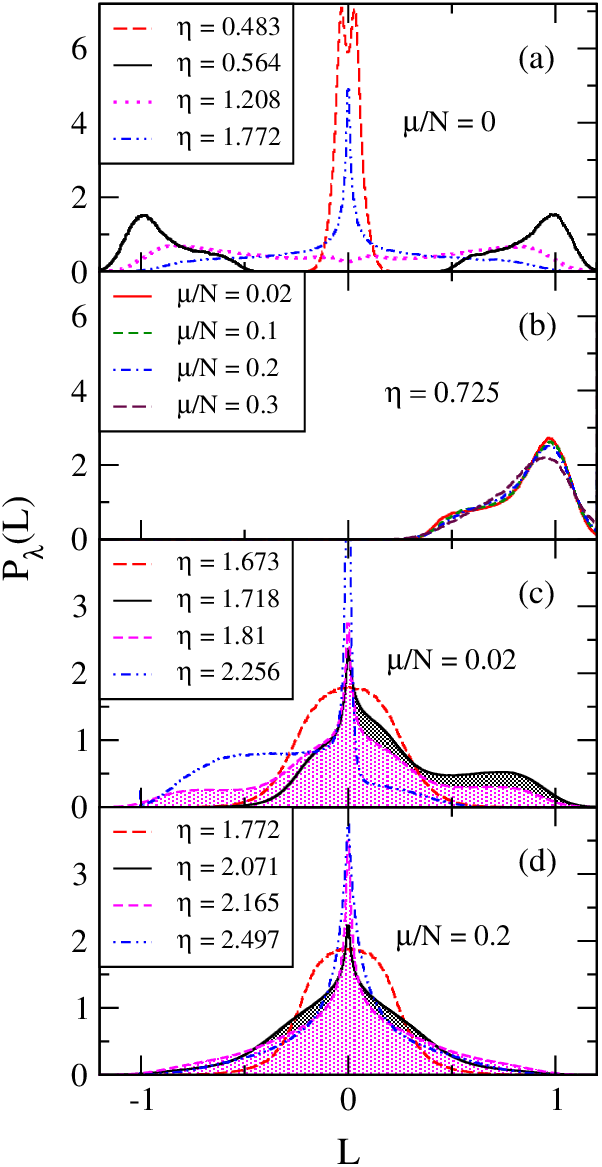}
\caption{(Color online) Probability distribution functions of the magnitude of the angular momentum of the individual particles $L_i$ for systems with $N=200$. In (a), for $\mu/N=0$, the two peaks away from the center for the curve with $\eta=0.564$ indicate a limit-cycle oscillatory state. That is also the case in (b), for the different values of $\mu/N$ with a one-peaked $P_{\lambda}(L)$ away from the center. In (c) and (d), the filled curves correspond to values of $\eta$ inside region III. In (c) the tails in some of the distributions indicate the development of limit-cycle-like oscillatory states. In contrast, in (d), the tails of the distributions decay faster from the center indicating the development of qualitatively disorder oscillatory states.}
\label{fig:mang-PDFs}
\end{figure}

Numerical results for different values of $\kappa$ (not shown) indicate that the system displays the same qualitative behavior, and the same stability regions, just as in the case considered here with $\kappa=1$. Moreover, the critical points that separate these regions behave similarly to the critical point $\eta_c$ for the case $\mu/N=0$, i.e., they shift to higher values as $\kappa$ increases while they shift to lower values as $\kappa$ decreases. Nonetheless, it was also observed that lager values of the connectivity $\mu/N$ are required for the phase transition from regions II to IV to become apparently continuous as $\kappa$ increases.

\subsection{\label{sec:angmom}Angular momentum}
Figures \ref{fig:mang}(a) and \ref{fig:mang}(b) show plots of the angular momentum $\Lambda$ as a function of $\eta$, with initial conditions taken in the TranS and the OscS, respectively. For $\mu/N=0$, $\Lambda$ fluctuates around zero for stationary translational states. On the other hand, for stationary oscillatory states, particles oscillate around the center of mass, on a limit-cycle, much in the same way as oscillatory states present in some other models \cite{shi96,lev00,erd05,dor06,rom08}. This can be appreciated in Fig.\ \ref{fig:mang-PDFs}(a) from the two peaks displayed by the stationary PDF, $P_{\lambda}(L)$, of the magnitude of the individual angular momenta $L_i=|{\bf L}_i|$ of the particles for $\eta=0.564$.

Having in mind that the OscS is always a solution for the system when $\mu/N=0$, becoming the only solution for $\eta>\eta_c$, the value of the order parameter $\Lambda$ will depend on initial conditions for noise values ($\eta\leq0.867$) to the left of the dash-dot-dashed vertical line in Figs.\ \ref{fig:mang}(a) and \ref{fig:mang}(b), whenever a stationary oscillatory state is reached. This can be understood from the fact that, in this region, and in the absence of alignment interactions, the correlated angular noise is not strong enough to randomly change the particles direction of rotation. In particular, in Fig.\ \ref{fig:mang}(b), initial conditions for the OscS were taken on the limit-cycle with half of the particles rotating in the clockwise direction, and half in opposite one, resulting in a $\Lambda=0$. Nonetheless, when initial conditions are taken in the TranS, but for noise values $\eta>\eta_c$ [to the right of the dotted vertical line in Fig.\ \ref{fig:mang}(a)], $\Lambda$ presents some bumps. Under these conditions, for values of the noise ($\eta_c<\eta\leq0.867$) between the dotted and dash-dot-dashed vertical lines in Fig.\ \ref{fig:mang}(a), the system undergoes a transition to the OscS after spending a transient time in the TranS (also reported in Ref.\ \onlinecite{erd05}). Thus, once the system reaches a stationary OscS, the number of particles rotating in each direction (clockwise and counterclockwise) may be asymmetric deriving in a fixed $\Lambda\geq0$. In contrast, for noise values ($\eta>0.867$) to the right of the dash-dot-dashed vertical line, where the correlated angular noise is strong enough to randomly change the particles direction of rotation in the OscS, $\Lambda\approx0$ regardless of initial conditions. However, for finite connectivities ($\mu/N>0$), $\Lambda$ shows the different stability regions identified before.

In region I, as the alignment interactions induce the majority of the particles to randomly select a direction of rotation, $\Lambda\approx1$ for the OscS. This state has a limit-cycle quality as $P_{\lambda}(L)$ shows in Fig.\ \ref{fig:mang-PDFs}(b). In region II only the TranS is accessible, thus, the value of $\Lambda$ is closer to zero.

\begin{figure}
\includegraphics[width=.48\textwidth]{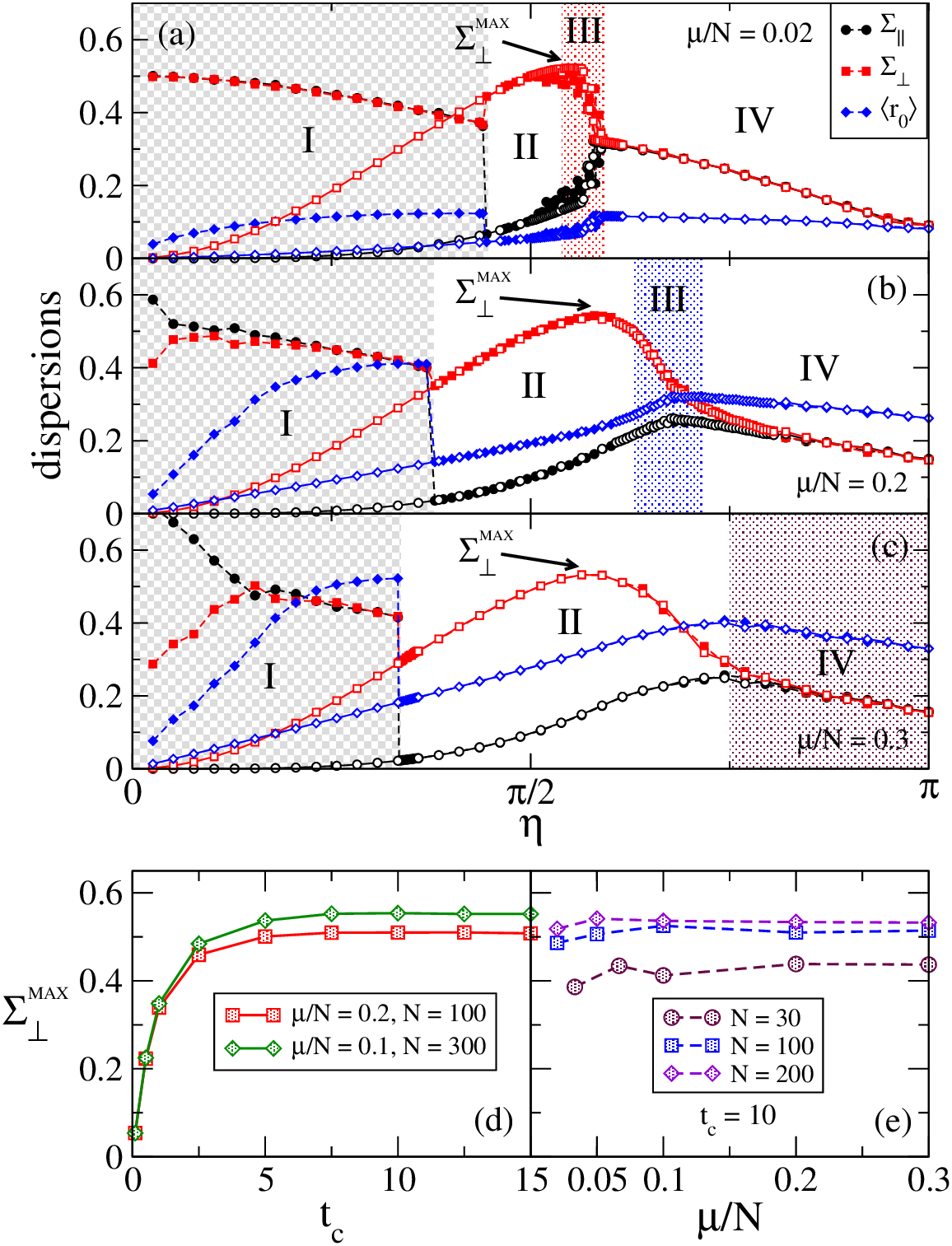}
\caption{(Color online) Above, (a)-(c), mean-square dispersions, given in Eqs.\ (\ref{eq:spar}) and (\ref{eq:sperp}), and the mean range of the alignment interaction, $\langle r_0 \rangle$ (see text for definition), as a function of $\eta$, for different connectivities, $N=200$ and $t_c=10$. The curves with dashed lines and solid symbols correspond to numerical results with initial condition in the OscS, while the curves with solid lines and clear symbols correspond to numerical results with initial condition in the TranS. The different stability regions are identified with roman numerals for the different cases. Below, behavior of the maximum transversal dispersion, $\Sigma_{\perp}^{\hbox{\tiny MAX}}$, as a function of (d) $t_c$, and (e) $\mu/N$.}
\label{fig:size}
\end{figure}

Inside region III, and for connectivities up to $\mu/N\approx0.2$, $\Lambda$ shows a discontinuous peak as the majority of the particles tend again to rotate in the same direction  whenever the system attains an OscS. This is corroborated by the tails of $P_{\lambda}(L)$ in Fig.\ \ref{fig:mang-PDFs}(c) for $\mu/N=0.02$; however, the direction of rotation is selected randomly considering that every OscS follows a free transition from the TranS. This type of OscS resembles a limit-cycle one even though the tails in $P_{\lambda}(L)$ are wider than in the previous cases due to particles that transit close to the center of mass [see Fig.\ \ref{fig:states}(b)]. The presence of a discontinuous peak in $\Lambda$ corresponds to cases where $P(\Psi)$ shows a bimodal distribution inside region III. Nonetheless, with the increasing connectivity, the amplitude of this peak decreases (see Fig.\ \ref{fig:mang}) along with the ``limit-cycle'' quality of the OscS.

\begin{figure}
\includegraphics[width=.42\textwidth]{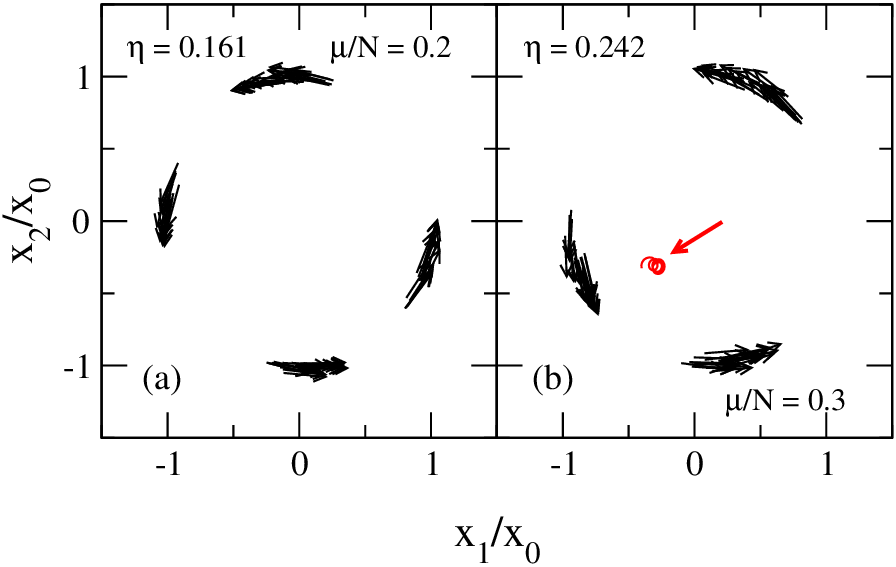}
\caption{(Color online) Snapshots of the configuration in position space for the model given in Eqs.\ (\ref{eq:first2}) and (\ref{eq:prop2}) with $\kappa = 1$, $t_c=10$, and $N=200$. The small arrows represent the velocity vectors ${\bf v}_i$ for each one of the particles. These states were obtained from simulations with initial conditions in the OscS. In (b) the large arrow points to the trajectory of the center of mass (solid line) for a finite period of time.}
\label{fig:states_cls}
\end{figure}

Of particular interest is the case with $\mu/N=0.2$ since, inside region III, it presents the discontinuous peak in $\Lambda$ along with a bimodal distribution for $P(\Psi)$; however, the distribution for $P_{\lambda}(L)$ looks rather symmetric throughout regions II to IV as shown in Fig.\ \ref{fig:mang-PDFs}(d). We will use this fact to define a crossover regime that separates the cases when the transition from region II to IV presents the bistable region III, from those when it seems continuous and region III is absent (see next subsection).

In region IV, the values of $\Lambda$ are closer to zero as the OscS becomes qualitatively disordered due to stronger noise, while $P_{\lambda}(L)$ does not show any tails but a symmetrical distribution around zero, in particular for $\mu/N\geq0.1$. For $\mu/N\geq0.3$, as region III disappears and the transition between the TranS and the OscS becomes apparently continuous, the discontinuous peak in $\Lambda$ also disappears, giving way to a smooth dependence of $\Lambda$ on $\eta$ instead. In this case, the OscS does not show any limit-cycle quality, looking rather disordered, while $P_{\lambda}(L)$ is always symmetrical throughout the transition, similar to the case for $\mu/N=0.2$. Thus, the apparently continuous transition from the TranS to the OscS, for connectivities $\mu/N\geq0.3$, may be considered an order-disorder.

\subsection{\label{sec:dispscal}Size of the system, dispersions, and critical points.}
In addition, the values of $t_c$ used in this work correspond to a regime where the size of the flock in position space, for any given $\eta$, does not depend on $t_c$ anymore. In consequence, the dependence of the critical points that separate the different stability regions on $t_c$ becomes also constant for a fixed values of $\kappa$ and $\mu/N$. This regime is defined by the condition $t_c\gg1$, and can be explained by looking at the behavior of dispersions (\ref{eq:disps}), shown in Figs.\ \ref{fig:size}(a) - \ref{fig:size}(c) as a function of $\eta$. In particular, by looking at the maximum of the transversal dispersion, $\Sigma_{\perp}^{\hbox{\tiny MAX}}$, and its dependence on $t_c$; as shown in Fig.\ \ref{fig:size}(d), it becomes constant for $t_c\gg1$. In the same limit, $\Sigma_{\perp}^{\hbox{\tiny MAX}}$ does not change much with $\mu/N$ but rather lightly with $N$ [see Fig.\ \ref{fig:size}(e)]. This means that the size of the system depends mainly on the noise intensity $\eta$ and the correlation time $t_c$, in the end becoming independent on $t_c$ for $t_c\gg1$. Thus, the critical points that separate the different stability regions will mainly depend on the ratio $\mu/N$ since the mean range of the alignment interaction, $\langle r_o \rangle(t)=\frac{1}{N}\sum_{i=1}^N[\frac{1}{\mu}\sum_{{\bf x}_j \in \, \zeta_i(t)} |{\bf x}_i(t) - {\bf x}_j(t)|]$, still depends on $\eta$ and $\mu/N$ itself, as can be appreciated in Figs.\ \ref{fig:size}(a) - \ref{fig:size}(c), and not on $t_c$ (in general, $\langle r_o \rangle$ is a function of all the parameters of the system). This is one of the main effects of the correlated angular noise, and allows an increase in the normalized connectivity $\mu/N$ to translate into an effective increase in the range of the alignment interaction, in position space, with respect to the overall size of the group for any given value of $\eta$ as shown in Figs.\ \ref{fig:size}(a) - \ref{fig:size}(c).

\begin{figure}
\includegraphics[width=.25\textwidth]{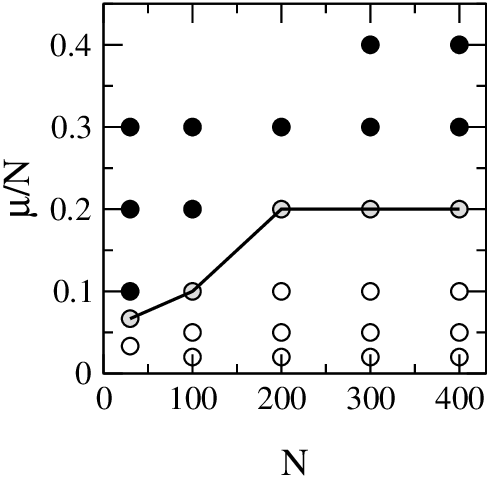}
\caption{Plot $\mu/N$ vs.\ $N$ where the solid line with grey circles corresponds to the crossover regime that separates the cases where the system exhibits a bistable region (region III) with free transitions between the TranS and the Oscs (depicted with clear circles), from those where the transition from the TranS to the OscS is apparently continuous (depicted with solid circles). In all of the cases $t_c=10$ and $\kappa=1$.}
\label{fig:scaling}
\end{figure}

We must remark that for cases with larger connectivities $\mu/N\geq0.2$, when initial conditions are taken in the OscS and for weak noises ($\eta\ll1$), the resulting stationary oscillatory states can present the formation of clusters as shown in Fig.\ \ref{fig:states_cls}. This comes as a result of the short-range character of the alignment interaction. Nonetheless, structures like these have been obtained by providing each particle with a short-range repulsive potential \cite{dor06}. On the other hand, the clusters, when present, seem to be very stable since the exchange of particles is not likely as they oscillate around the center of mass in a limit-cycle fashion. The segregation of the system in this way is the main cause that the dispersions parallel ($\Sigma_{||}$) and orthogonal ($\Sigma_{\perp}$), given in Eqs.\ (\ref{eq:disps}), are different in Figs.\ \ref{fig:size}(b) and \ref{fig:size}(c) for weak noise intensities. What happens is that, since the particles are not homogeneously distributed in a circular form around the center of mass, the center of mass itself does not lie in the center of the circular formation but closer to the largest cluster. In consequence, the center of mass is able to perform a translational motion rather than a diffusive one. This is shown in Fig.\ \ref{fig:states_cls}(b), where the large arrow points to the trajectory of the center of mass for a finite time. The exchange of particles among the clusters is more likely when noise increases, in the end leading to the destruction of these structures.

Finally, Figure \ref{fig:scaling} shows a plot of $\mu/N$ vs.\ $N$, and is intended as a scaling analysis of the transition from regions II to IV. One can observe that as the number of particles $N$ in the system increases, the crossover regime (solid line with grey circles) that separates the cases when the bistable region III is absent, and the transition from the TranS to the OscS seems to be continuous with a one-peaked $P(\Psi)$ at criticality (solid black circles above the line), from those when a region III is present and $P(\Psi)$ is bimodal (clear circles below the line), moves from lower values of the connectivity $\mu/N$ to larger ones, finally stabilizing for $N\geq200$. As far as our numerical results indicate, increasing $N$ any further does not change the position of the different stability regions for finite values of $\mu/N$ and $t_c\gg1$.

\section{\label{sec:concs}Conclusions}
\begin{figure}
\includegraphics[width=.35\textwidth]{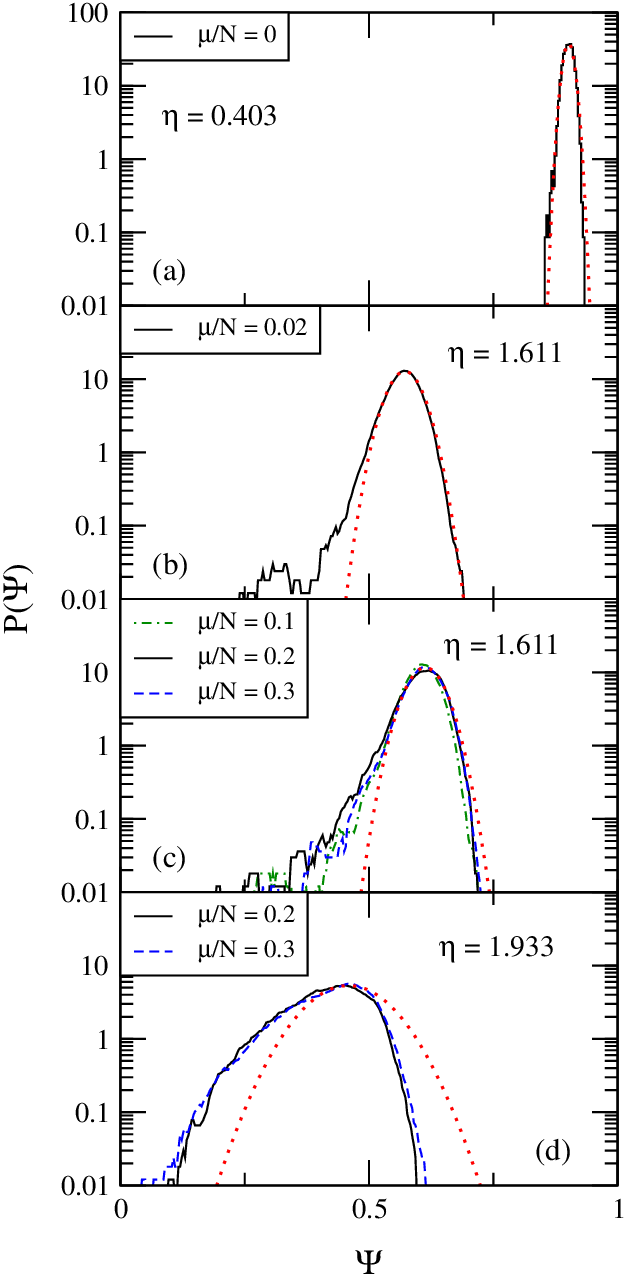}
\caption{(Color online) [(a)-(d)] Semi-log plots of the PDF of $\Psi$ with $N=200$,  $t_c=10$, and $\kappa=1$, for different values $\mu/N$ and $\eta$. In (a), the selected value of $\eta$ lies in the subcritical region of the phase transition from the TranS to the OscS. In (b)-(d), the selected values of $\eta$ lie inside region II. The dotted lines correspond to Gaussian fits of the whole distribution in (a), and of the crest of the distributions in (b)-(d).}
\label{fig:interm}
\end{figure}

We have studied the mixed effects of correlated angular noise, centering and alignment interactions in a simple two-dimensional (2D) Lagrangian model of $N$ self-propelled particles. As a result, we have observed rich dynamics in the collective behavior along with different stability regions as the amplitude of the noise changes. Of particular interest is the development of intermittent behavior as the presence of alignment interactions induce turbulence bursts in the TranS where the group is able to even segregate into clusters. During these events, the group may present local order but not global, and for certain combinations of the parameters, they allow the system to freely transit from the TranS to the OscS and vice versa. This result is \emph{distinctive} from other models where the collective dynamics is condemned to live in monotonous stable states. Indeed, a closer inspection of $P(\Psi)$ shows the presence of a tail that deviates from a Gaussian distribution and accounts for these turbulence bursts inside region II, while this behavior is not observed in the case $\mu/N=0$ [see Figs.\ \ref{fig:interm}(a) - \ref{fig:interm}(c)]. Moreover, increasing the range of the alignment interaction (by increasing $\mu/N$) enhances this effect that finally leads to a fundamental change in the dynamics of the system as the bistable region III disappears, and the transition in $\Psi$ from the TranS of region II to the OscS of region IV apparently changes its nature to continuous [see Figs.\ \ref{fig:psi}(a) and \ref{fig:scaling}]. This comes about from the fact that, for large enough connectivities ($\mu/N\geq0.2$) and subcritical values of $\eta$ closer to the end of region II, the system becomes fully intermittent as $P(\Psi)$ shows a non-Gaussian asymmetrical form [see Fig.\ \ref{fig:interm}(d)], consequently leading to the existence of region III and its further disappearance when the connectivity increases. This strongly suggests what seems to be a clear connection between the intermittent behavior in the collective motion of the group and the kind of phase transitions undergone by its mean velocity.

The kind of dynamics the system develops inside region II, resembles that in the model of Vicsek \emph{et al.}\ \cite{vic95}, where the cluster formation has proven to play an essential role in the development of intermittent dynamics \cite{hue04,cha08}. This comes accompanied by a continuous order-disorder phase transition in the mean velocity of the group \cite{per08a,ton95,nag07,bag08,natph}. The intermittent behavior shows up as, in the stationary ordered phase, a laminar regime (where all the particles move in a rather ordered fashion) is interrupted by chaotic bursts of turbulence where the system segregates in clusters that may present local order, but not global, in a very similar way as our model does. Moreover, the authors in Ref.\ \cite{rom08} reported the cluster formation in a Lagrangian model that considers alignment forces, centering, and separation, but with white additive noise. In fact, intermittent behavior can be observed in other systems far from equilibrium including biological ones (see references in \cite{hue04}).

Finally, as our numerical simulations show, the different dynamics displayed by our model seem to be robust even if each particle is provided with a hard-core repulsive potential. This case is analyzed in detail in \cite{dos08}. On the other hand, the nonlinear way angular noise enters in our model results in the development of an effective individual mean speed that depends on $\eta$, and in contrast to some other models where it does not \cite{lev00,erd05,dor06,lee06,rom08}. It can be shown that $v_i$ fluctuates around the value $\frac{v_0}{\eta}\sin\eta$, since $\langle v_i \rangle$ depends only on the form of the distribution of the noise. We believe our results may be relevant in the theory of flocking and, in general, in the theory of phase transitions in systems out of equilibrium.

\begin{acknowledgments}
The authors are grateful to M. Aldana (UNAM, Mexico) for fruitful discussions, to the Center for High Performance Computing at UNM, and to the NMCAC for the use of their facilities. This work was partially supported by the NSF under Grant No. INT-0336343.
\end{acknowledgments}

\appendix

\section{\label{sec:limitcases}Limit cases}

In this appendix we present the analysis of some limit cases of our model, and of other well-known models, as we compare them in the light of some analytical estimations.

First, let us consider the global alignment interaction limit of model (\ref{eq:first2}), where the individual propulsive direction $\psi_i$ of each particle, given in Eq.\ (\ref{eq:prop2}), coincides with the direction $\psi_{CM}$ of the center of mass velocity ${\bf V}_{CM}$, i.e., $\psi_i=\psi_{CM}$ at all times. This case can be solved exactly for the order parameter $\Psi$, and may arise from taking $\mu=N-1$. The only solution that the system presents is that one for the TranS regardless of the initial conditions, and can be calculated by solving the equation of motion for the center-of-mass. The latter is obtained by summing Eqs.\ (\ref{eq:first2}) and dividing by the number of particles $N$,
 \begin{eqnarray}
\lefteqn{m\frac{d{\bf V}_{CM}}{dt} = - \frac{\gamma}{v_0}{\bf V}_{CM}} \nonumber \\
& & + \frac{\gamma}{N}\sum_{i=1}^{N}[\cos(\psi_{i}+\phi_{i}){\bf i}+\sin(\psi_{i}+\phi_{i}){\bf j}],
\label{vcm}
\end{eqnarray}
where the effects of angular noise are captured in the expression $\cos(\psi_{i}+\phi_{i}){\bf i}+\sin(\psi_{i}+\phi_{i}){\bf j},$ with $\psi_{i}$ and $\phi_{i}$ defined as before.

We have then, Eq.\ (\ref{vcm}) is akin to the Langevin equation $m\frac{d{\bf V}_{CM}}{dt} = -\frac{\gamma}{v_0}{\bf V}_{CM} + \xi(t)$, where the random force $\xi$ is defined as the random quantity $\frac{\gamma}{N}\sum_{i=1}^{N}[\cos(\psi_{i}+\phi_{i}){\bf i}+\sin(\psi_{i}+\phi_{i}){\bf j}]$, and can be interpreted as a directed random walk with persistence \cite{per07}. Equation (\ref{vcm}) may be split into two coupled equations for the magnitude $V_{CM}=|{\bf V}_{CM}|$ and  the direction $\psi_{CM}\equiv\arctan\left[\sum_{i}^{N}v_{y,i}/\sum_{i}^{N} v_{x,i}\right] $ of the center-of-mass velocity. After taking an ensemble averages over different realizations of the stochastic variables $\phi_i$, the equations may be explicitly written as
\begin{subequations}
\label{vcmsplit}
\begin{equation}
\frac{d\langle V_{CM}\rangle}{dt}=-\frac{\gamma}{mv_{0}}\langle V_{CM}\rangle +\frac{\gamma}{mN}\sum_{i=1}^{N}\langle\cos\bar{\phi}_i\rangle,
\label{vcmmag}
\end{equation}
and
\begin{equation}
\left\langle V _{CM}\frac{d\psi_{CM}}{dt}\right\rangle =\frac{\gamma}{mN}\sum_{i=1}^{N}
\langle\sin\bar{\phi}_i\rangle,
\label{vcmdir}
\end{equation}
\end{subequations}
respectively, where $\bar{\phi}_i\equiv\psi_{i}-\psi_{CM}+\phi_{i}$ with a PDF $P'(\bar{\phi})$ equivalent for all of the particles.

In particular, for the global alignment interaction case, equations (\ref{vcmsplit}) may be solved straightforwardly since $\psi_{i}=\psi_{CM}$ at all times, and $\bar{\phi}_i=\phi_{i}$ (thus \mbox{$P'(\bar{\phi})=P_{\eta}(\phi)$}, the latter corresponding to the distribution of the noise). Then, the rightmost terms reduce to $\langle\cos\bar{\phi}_i\rangle=\langle\cos\phi_{i}\rangle$ and $\langle\sin\bar{\phi}_i\rangle=\langle\sin\phi_{i}\rangle$.

\begin{figure}
\includegraphics[width=.48\textwidth]{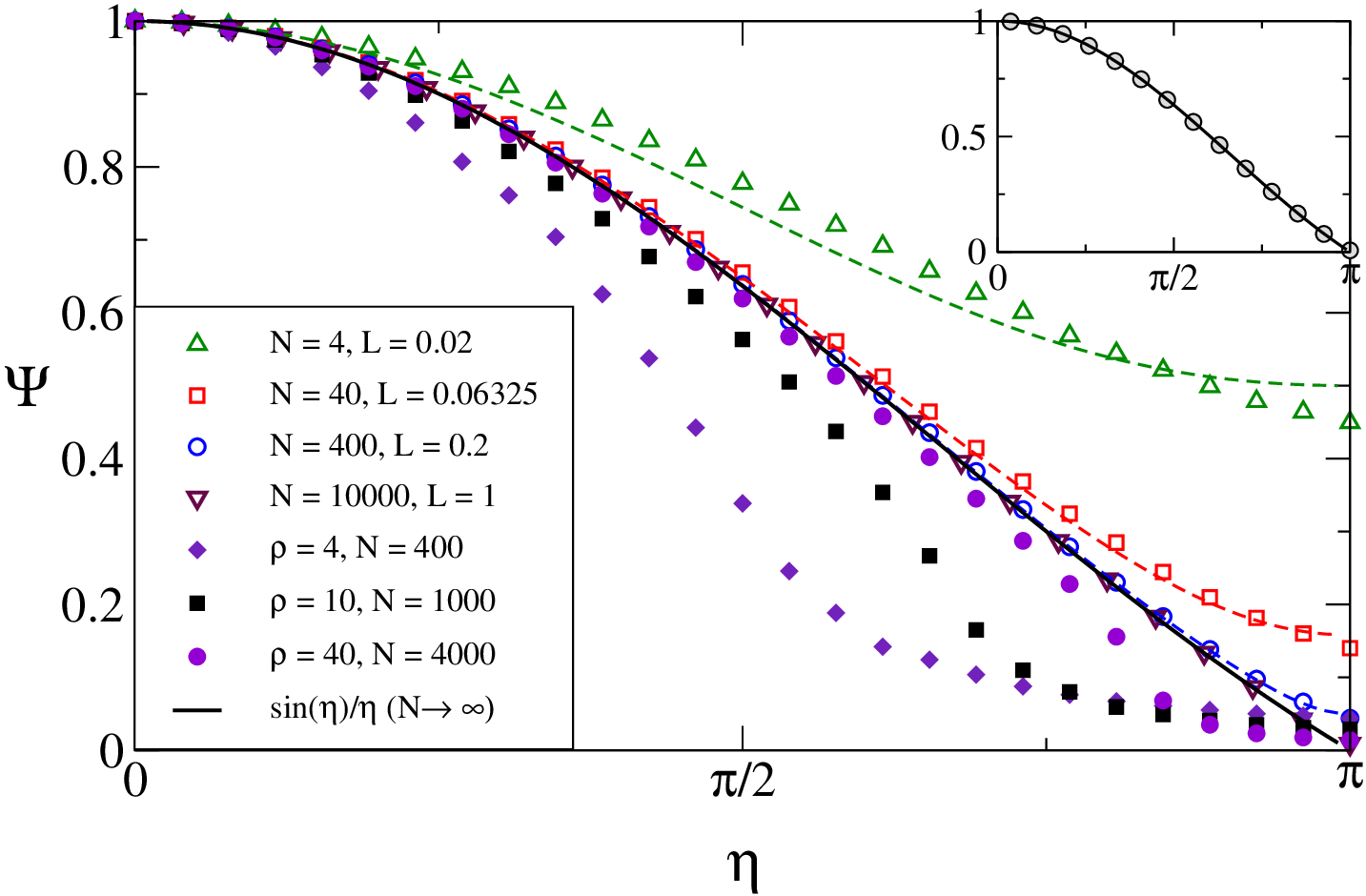}
\caption{(Color online) Normalized magnitude of the mean velocity of the group, in the steady state, given by the order parameter $\Psi$ of Eq.\ (\ref{Psi0}), for the Vicsek \emph{et al.}\ model, as a function of the noise intensity $\eta$ while the limit $N\rightarrow\infty$ is approached. Numerical (curves with clear symbols) and analytical (dashed lines) results, the latter corresponding to the expression of Eq.\ (\ref{eq:Vic_global}) for different values of $N$, are presented assuming a global alignment interaction and a constant density $\rho=10^3$. On the other hand, when a local alignment interaction is assumed, numerical results (curves with solid symbols) are presented as $N\rightarrow\infty$ by making $\rho\rightarrow\infty$ for a fixed $L=10$. The black solid curve corresponds to the analytical approximation of Eq.\ (\ref{eq:Vic_inf-dens}). In all the simulations for the Vicsek \emph{et al.} model only the \emph{small velocity regime} \cite{nag07} was considered ($v_0=0.03$) with $r=1$. The inset shows numerical results (grey circles) for model (\ref{eq:first2}) with $N=200$, $k=1$, and $t_c=10$, when a global alignment interaction ($\mu/N=N-1$) is considered, against the analytical exact result (black solid curve) of Eq.\ (\ref{eq:tmspsiglob}).}
\label{fig:psi_vskglob}
\end{figure}

By definition the probability distribution of the noise, $P_{\eta}(\phi)$, given in Eq.\ (\ref{eq:peta}), is the same for all the angles $\phi_{i}$; therefore the quantities $\langle\cos\phi_{i}\rangle$  and $\langle\sin\phi_{i}\rangle$ do not depend no the sub-index $i$, and have the values $\frac{1}{\eta}\sin\eta$ and  $0$, respectively. The solution to Eq.\ (\ref{vcmmag}) is then given by
\begin{eqnarray}
\lefteqn{\langle V_{CM}(t)\rangle =} \nonumber \\ 
& & \langle V_{CM}(0)\rangle \, e^{-\frac{\gamma}{mv_{0}}t}
+ \frac{v_{0}}{\eta}\sin\eta\left[1-e^{-\frac{\gamma}{mv_{0}}t}\right].
\label{vcmmags}
\end{eqnarray}
Under the assumption that the system is \emph{ergodic}, we will replace the temporal average in Eq.\ (\ref{eq:vcm1}) by an ensemble average in the stationary state $\langle V_{CM}^{st}\rangle$ which, from expression (\ref{vcmmags}), has the value $v_{0}\frac{\sin\eta}{\eta}$. Finally, we get the following exact result for the order parameter:
\begin{equation}
\Psi=\frac{1}{\eta}\sin\eta.
\label{eq:tmspsiglob}
\end{equation}
This expression is compared to results obtained from numerical simulations in the inset of Fig.\ \ref{fig:psi_vskglob}. The agreement between theory and numerical data is extremely good, which also validates the numerical integration scheme used in this work. Notice how this solution only depends on the noise intensity $\eta$.

It is interesting to note that the order parameter in the Vicsek \emph{et al.} model \cite{vic95}, when a global alignment interaction is considered in the thermodynamic limit, i.e., when $N\rightarrow\infty$ for a constant density $\rho=N/L^2$, shows exactly the same dependence on $\eta$ as that of Eq.\ (\ref{eq:tmspsiglob}). Indeed, in Ref.\ \onlinecite{vic95}, the instantaneous order parameter $\Psi(t)\equiv\vert\bm{V}_{CM}(t)\vert /v_{0}$ is also a measure of the coherent behavior of the system and by definition given by
\begin{eqnarray}
\Psi(t+\Delta t) & = & \frac{1}{N}\left[\left(\sum_{i=1}^{N}\sin\varphi_{i}(t+\Delta t )\right)^{2} \right.
\nonumber \\
& & \left. + \left(\sum_{i=1}^{N} \cos\varphi_{i}(t+\Delta t)\right)^{2}\right]^{1/2},
\label{Psi0}
\end{eqnarray}
where the phases $\varphi_i$ correspond to the directions of the individual velocities of $N$ particles contained in a box of size $L$ with periodic boundary conditions. The system considers local alignment interactions (see below), while the velocities ${\bf v}_i$ of the particles are determined simultaneously at each time step. Then, their positions are updated through
\begin{subequations}
\label{eq:vicrule}
\begin{equation}
{\bf x}_i (t + \Delta t) = {\bf x}_i (t) + {\bf v}_i (t) \Delta t.
\label{eq:vicpos}
\end{equation}
Here, the velocity was constructed to have a fixed magnitude $v_0$ equal for all of the particles with $\Delta t = 1$. In this way, the particles advance the same fixed distance in the direction of the angle $\varphi_i (t + \Delta t)$ at each time step, the latter calculated from
\begin{equation}
\varphi_i (t + \Delta t) = \langle \varphi_i (t) \rangle_r + \phi_i (t),
\label{eq:vicph}
\end{equation}
\end{subequations}
where $\langle \varphi_i (t) \rangle_r$ denotes the mean direction of the velocities of the particles (including particle $i$) that lie within a circle of radius $r$ surrounding a given particle, while $\phi_i$ is a random angle taken from a flat distribution between $[-\eta, \eta]$ (here defined in this way to be able to compare with our model), with $\eta$ running from 0 to $\pi$ \cite{nag07}. 

It is possible to give an expression for $\Psi$ only in terms of the random quantities $\phi_{i}$, i.e., by assuming a global alignment interaction such that $\langle \varphi_i (t) \rangle_r = \psi(t)$, where $\psi(t)$ is the same for all of the particles and corresponds to the mean direction of the whole group. This can be achieved by making the range of the alignment interaction larger or, at least, equal to the size of the system ($r \ge L$). After taking the sine and cosine of Eq.\ (\ref{eq:vicph}), and summing over all particles, one gets
\begin{eqnarray*}
\sum_{i=1}^{N}\sin\varphi_{i}(t+\Delta t)&=& \sin\psi(t)\sum_{i=1}^{N}\cos\phi_{i}(t) \\
&&+ \cos\psi(t)\sum_{i=1}^{N}\sin\phi_{i}(t),
\end{eqnarray*}
and
\begin{eqnarray*}
\sum_{i=1}^{N}\cos\varphi_{i}(t+\Delta t)&=&\cos\psi(t)\sum_{i=1}^{N}\cos\phi_{i}(t) \\
&&- \sin\psi(t)\sum_{i=1}^{N}\sin\phi_{i}(t).
\end{eqnarray*}
By substituting last expressions in Eq.\ (\ref{Psi0}), expanding the squares, and simplifying terms one can write
\begin{eqnarray}
\Psi(t+\Delta  t)=\frac{1}{N}\left[ \left(\sum_{i=1}^{N} \cos\phi_{i}(t)\right)^{2} \right. \nonumber \\
\left. +\left(\sum_{i=1}^{N}\sin\phi_{i}(t)\right)^{2}\right]^{1/2}.
\end{eqnarray}
An evaluation of $\langle\Psi(t+\Delta  t)\rangle$ in this form is difficult. Instead, let us compute the quantity $N^{2}\langle\Psi^{2}(t+\Delta  t)\rangle$,
\begin{eqnarray*}
N^{2}\langle\Psi^{2}(t+\Delta  t)\rangle&=&\left\langle\left(\sum_{i=1}^{N} \cos\phi_{i}(t)\right)^{2}
\right. \nonumber \\ 
& & + \left. \left(\sum_{i=1}^{N} \sin\phi_{i}(t)\right)^{2}\right\rangle
\end{eqnarray*}
Making use of the \emph{multinomial theorem} \cite{abr72}, after some algebra one gets:
\begin{eqnarray*}
N^{2}\langle \Psi^{2} \rangle & = & 
N+2\frac{\sin^{2}\eta}{\eta^{2}}\sum_{i=1}^{N-1}\sum_{j>i}^{N}1\\
& = & N + N(N-1)\frac{\sin^{2}\eta}{\eta^{2}},
\end{eqnarray*}
and thus
\begin{equation}
\sqrt{\langle \Psi^{2} \rangle} = 
\sqrt{\frac{1}{N}+\left(1-\frac{1}{N}\right)\frac{\sin^{2}\eta}{\eta^{2}}}.
\label{eq:Vic_global}
\end{equation}
In the limit when $N\longrightarrow\infty$, and on the basis of some observations from our numerical results, one can write
\begin{equation}
\Psi\approx\sqrt{\langle \Psi^{2} \rangle}=\frac{\sin\eta}{\eta}.
\label{eq:Vic_inf-dens}
\end{equation}
The last approximation corresponds to the asymptotic behavior, in the steady state, for the order parameter of the Vicsek \emph{et al.} model \cite{vic95} under the assumptions of a global alignment interaction in the thermodynamic limit ($N\rightarrow\infty$). Under the same global alignment assumption, the expression of Eq.\ (\ref{eq:Vic_inf-dens}) has also been derived by approximating the system with a directed random walk with persistence \cite{per08b}.

Figure \ref{fig:psi_vskglob} shows plots of results obtained from numerical simulations (curves with clear symbols) for the behavior of $\Psi(\eta)$ against the analytical expression of Eq.\ (\ref{eq:Vic_global}) (dashed curves), considering a constant density $\rho=10^3$, and system sizes such that $L \le r$ as $N$ tends to infinity. As apparent from the figure, the order parameter approaches the expression of Eq.\ (\ref{eq:Vic_inf-dens}) from above, and the absence of the phase transition is evident. For all the numerical simulations of the Vicsek \emph{et al.} model \cite{vic95}, we have set $\Delta t=1$, $r=1$, and $v_0=0.03$, the latter well inside the \emph{small velocity regime} ($v_0 \le 0.1$) \cite{nag07}.

In this regime, particles that were within the same interaction vicinity at time $t$ will most likely remain within the same interaction vicinity at time $t + \Delta t$. Nonetheless, in the computation of Eq.\ (\ref{eq:Vic_inf-dens}) we have made no assumptions on the particles speed $v_0$, nor the linear size of the system $L$. Moreover, the same expression has been obtained for the vectorial network model (VNM) of Aldana \emph{et al.}\ \cite{ald03}, in Eq.\ (23b) of Ref.\ \onlinecite{pim08}, when only \emph{intrinsic} (or angular) noise is considered for an infinite network connectivity. This case corresponds to the global alignment interaction limit in both: our model, given in Eq.\ (\ref{eq:first2}), and the Vicsek \emph{et al.}\ model. The VNM of Aldana \emph{et al.}, as stated there, may be considered as a mean-field approximation of the Vicsek \emph{et al.}\ model in the limit of \emph{extreme particle speeds} ($v_0 \sim 1000$ for simulation purposes) \cite{nag07}; indeed, when $v_0\rightarrow\infty$. One of the conditions for the validity of this mean-field approximation is that spatial correlations developed due to the motion of the particles, and present in the small velocity regime, are destroyed at every time step when the particles are allowed to move with extreme speeds. However, as mentioned before, our calculation on the Vicsek \emph{et al.}\ model does not assume a particular regime for $v_0$.

The common behavior displayed by $\Psi(\eta)$ in these three models, is only due to the presence of the global alignment interaction in combination with angular noise, which in fact are the only features they share. This implies that the global alignment interaction aids to the homogenization of the systems, as it dominates over any other characteristic dynamics of a given particular model, inducing an effective mean field for all of the interacting particles.

On the other hand, when local alignment interactions ($r<L$) are assumed for the Vicsek \emph{et al.}\ model, it is well known that the critical point $\eta_c$ (that depends on $\rho$) moves to higher values as the density is increased, finally reaching the ``infinite temperature'' limit, i.e., $\eta_c\rightarrow\pi$ as $\rho\rightarrow\infty$, where the phase transition disappears \cite{vic95}. For this case, and with fixed $L=10$, our numerical simulations show that the order parameter approaches the expression given in Eq.\ (\ref{eq:Vic_inf-dens}) from below (curves with solid symbols in Fig.\ \ref{fig:psi_vskglob}). Notice how the critical point $\eta_c$ of the phase transition is shifted to higher values with the increasing $\rho$. As density tends to infinity, an thus $N\rightarrow\infty$, in the thermodynamic limit, it is plausible to expect the fulfillment of the condition for the global alignment interaction [$\langle \varphi_i (t) \rangle_r = \psi(t)$] once more, thus, leading to the same result as the one presented in Eq.\ (\ref{eq:Vic_inf-dens}).

This may be explained from the combination of two different effects. On the one hand, because of the presence of noise in the system, particles move randomly with some correlation due to the local alignment interactions. As density is increased, this ``random'' motion makes the system spatially more homogeneous, in contrast to the low-density case where particles tend to form groups that move coherently in random directions \cite{hue04}. Then, the overlap of the interaction vicinities of the different particles increases, and the system becomes more coherent developing the effective mean-field interaction previously discussed for the global alignment interaction limit.



\begin{thebibliography}{99}
\bibitem{natph} The authors in \cite{gre03,gre04,cha08} claim that the phase transition is discontinuous even in the presence of additive angular noise.
\end{thebibliography}
\end{document}